\documentclass[conference]{IEEEtran}
\IEEEoverridecommandlockouts
\usepackage{booktabs}
\usepackage{array} 
\usepackage{tabularx}
\usepackage{multirow}
\usepackage{graphicx}  
\usepackage{float}  
\usepackage{subfigure}  
\usepackage{cite}
\usepackage{amsmath,amssymb,amsfonts}
\usepackage{algorithmic}
\usepackage{textcomp}
\usepackage{xcolor}
\usepackage{lettrine}
\def\BibTeX{{\rm B\kern-.05em{\sc i\kern-.025em b}\kern-.08em
    T\kern-.1667em\lower.7ex\hbox{E}\kern-.125emX}}

\begin{document}

\title{KGAMC: A Novel Knowledge Graph Driven Automatic Modulation Classification Scheme\\
\thanks{This work was supported by National Key R\&D Program of China under Grant 2023YFB2904500, the National Natural Science Foundation of China under Grant 62222107, Grant 62071223, the Postgraduate Research \& Practice Innovation Program of Jiangsu Province SJCX23\_0100 and Zhejiang Lab Open Research Project under Grant K2022PD0AB09. }
}

\author{ Yike Li$^{ \S }$, Lu Yuan$^{ \S }$, Fuhui Zhou$^{ \S\P }$, Qihui Wu$^{ \S }$, Naofal Al-Dhahir$^{ \ddag }$, and Kai-Kit Wong$^{\dagger}$\\
 $^{\S}$Nanjing University of Aeronautics and Astronautics, China, $^{\P}$Zhejiang Lab, China\\
 $^{\ddag}$The University of Texas at Dallas, USA,
 $^{\dagger}$University College London, London, UK\\
Email: \emph{liyikederek, yuanlu@nuaa.edu.cn, zhoufuhui@ieee.org, wuqihui2014@sina.com,}\\
\emph{aldhahir@utdallas.edu, and kai-kit.wong@ucl.ac.uk}
}

\maketitle

\begin{abstract}
Automatic modulation classification (AMC) is a promising technology to realize intelligent wireless communications in the sixth generation (6G) wireless communication networks. Recently, many data-and-knowledge dual-driven AMC schemes have achieved high accuracy. However, most of these schemes focus on generating additional prior knowledge or features of blind signals, which consumes longer computation time and ignores the interpretability of the model learning process. To solve these problems, we propose a novel knowledge graph (KG) driven AMC (KGAMC) scheme by training the networks under the guidance of domain knowledge. A modulation knowledge graph (MKG) with the knowledge of modulation technical characteristics and application scenarios is constructed and a relation-graph convolution network (RGCN) is designed to extract knowledge of the MKG. This knowledge is utilized to facilitate the signal features separation of the data-oriented model by implementing a specialized feature aggregation method. Simulation results demonstrate that KGAMC achieves superior classification performance compared to other benchmark schemes, especially in the low signal-to-noise ratio (SNR) range. Furthermore, the signal features of the high-order modulation are more discriminative, thus reducing the confusion between similar signals.

\end{abstract}

\section{Introduction}
\lettrine{\textbf{A}}{\textbf{utomatic}} modulation classification (AMC) plays an important role in the intelligent wireless communications and is one of the significant components of intelligent spectrum management \cite{b16}. In civilian dynamic spectrum access technology, AMC can detect the modulation used by unauthorized users. In military spectrum management, AMC facilitates the detection of abnormal users, signal analysis and the realization of decryption, and detecting enemy signals. With the in-depth study of 6G wireless communication networks, dynamic and complex wireless environment and explosive growth of wireless terminal devices impose new challenges to the modulation recognition for 6G wireless communication systems, and thus raise urgent need to propose novel AMC schemes \cite{b1}.

 Traditional AMC schemes, including the maximum likelihood method on hypothesis testing and the feature extraction method based on expert knowledge, are complex, time-consuming, and cannot adapt to the dynamic wireless environment \cite{b3}. With the rapid development of artificial intelligence, the deep learning (DL) methods are widely applied to improve the recognition efficiency and robustness \cite{b2}. The existing DL methods can be mainly classified into two categories, namely, data-driven AMC and data-and-knowledge dual-driven AMC. The data-driven AMC methods employ the DL networks to learn signal feature extraction pattern from the distributions of massive signal samples, such as deep neural networks (DNN), convolutional neural network (CNN), recurrent neural network (RNN) and so forth \cite{b3}. These networks are proved to be effective in other classification tasks. Nevertheless, they still encounter the data-dependency problem and lack of classification interpretability, which render classification results unreliable especially in a complex electromagnetic environment. To address these challenges, the data-and-knowledge dual-driven AMC methods combine the data features with expert knowledge to offer more information about the data, enhancing the performance and the interpretability of data-driven networks. The related works are categorized as follows.

\textit{Data-driven AMC methods:} O'Shea \textit{et al}. \cite{b4} were the first to use CNN as a data-driven DL model for signal feature extraction to confirm the effectiveness of DL models in AMC.  A long short-term memory (LSTM) was later adopted by the authors in \cite{b5} to obtain the sequence characteristics of signals in the time domain and achieve high classification accuracy. However, it requires a long training time due to its recurrent structure. Both of the methods above only focus on the spatial features of the signal waveform or the temporal features of the signal sequence. To utilize both features of the modulated signals, a combination of CNN and gated recurrent unit (GRU) model named PET-CGDNN was proposed in \cite{b6}, which achieves impressive modulation classification accuracy.

\textit{Data-and-knowledge dual-driven AMC methods:} The authors in \cite{b7} utilized two pre-trained models with different training objects, namely, visual model and attribution model. The data features are extracted by the visual model and the expert knowledge of the signal is learnt by the attribution model. Then, the attribution knowledge was converted to the data features space to aggregate visual vectors for enhancing the AMC performance. To further enhance the performance of the data-and-knowledge dual-driven AMC methods, a hybrid knowledge-and-data dual-driven DL framework in \cite{b8} was proposed. The authors provided each signal sample with its own handcraft expert features, and designed a data-knowledge features fusion mechanism with an attention layer. Furthermore, to improve the universality of the models, the authors of \cite{b9} fine-tuned the distribution of data features in high-dimensional vector space with expert knowledge to address the open set recognition problem in modulation recognition.

However, the data-and-knowledge dual-driven AMC methods discussed above failed to reduce the computational complexity due to the additional prior knowledge computation required for each blind signal, and neglected the opportunity to further boost the model performance by guiding the learning of model with knowledge. To address these issues, we investigate a new application of knowledge within the data-and-knowledge AMC scheme. Knowledge graph (KG) has been demonstrated to be an effective approach for representing the knowledge and the knowledge hierarchy. Recently in \cite{b10}, KG has been applied as a fundamental knowledge database in the contrastive learning of chemical molecular property prediction, showing extraordinary ability in enhancing the interpretability and performance of the feature extractor. 

Motivated by the KG potential in enhancing the performance of other data-driven models, a novel knowledge graph driven AMC (KGAMC) scheme is proposed in this paper. It is the first time that a modulation KG (MKG) is constructed to enhance the interpretability of the model classification mechanism. Moreover, an improved relation-graph convolution network (RGCN) with a residual connection branch is designed to extract the semantic features. Based on the constructed MKG and RGCN, a novel knowledge graph driven framework for AMC is proposed to leverage the semantic difference of knowledge, enabling the data-driven model to learn a new feature extraction pattern. Simulation results exhibit compelling evidence of the superior classification performance of our proposed KGAMC compared to other DL-based AMC schemes, particularly at low SNR. Furthermore, the signal features of the high-order modulations are more discriminative, reducing the confusion between the similar signals.

The remainder of this paper is organized as follows. The preliminaries are presented in Section II. Section III presents our proposed scheme. Section IV presents the simulation results. Finally, the paper concludes with Section V.

\section{Preliminaries}
\subsection{Signal Model}
Generally, in the modulation classification problem, the received signal $r(n)$  can be expressed as
\begin{equation}
r(n)=s(n)+w(n), n=1,2,...,L,\
\end{equation}
where $L$ is the length of the discrete-time series, $s(n)$ denotes the  $n$-th (complex) symbol, and $w(n)$ is the additive white Gaussian noise (AWGN) with zero mean and $\delta_{w}^{2}$ variance.

In DL-based AMC, the original discrete signals are typically preprocessed into an I/Q vector, since those two parts usually obey an identical distribution \cite{b11}. And the I/Q samples can be expressed as a vector, given as
\begin{subequations}
\begin{align}
  {{\mathbf{r}}_{i}}&={{\mathbf{I}}_{i}}\mathbf{+}{{\mathbf{Q}}_{i}}, \\ 
 &=\Re ({{\mathbf{r}}_{i}})+j\Im ({{\mathbf{r}}_{i}}),  
\end{align}
\end{subequations}
where ${\mathbf{I}}_{i}$ and ${\mathbf{Q}}_{i}$ denote the in-phase and the quadrature parts of the received signal, respectively, and $j$ is the unit imaginary number. $\Re ({{\mathbf{r}}_{i}})$ and $\Im ({{\mathbf{r}}_{i}}) $ represent the operators of real and imaginary parts of the signal, respectively. The raw data can be expressed as\
\begin{equation}
    {{\mathbf{r}}_{i}}=\left( \begin{matrix}
       \Re [r(1),r(2),...,r(L)]  \\
       \Im [r(1),r(2),...,r(L)]  \\
    \end{matrix} \right).
\end{equation}
\subsection{Problem Formulation}
In the knowledge graph driven AMC scheme, two feature extractors $f_{\theta},g_{\sigma}$ and a classifier $h_{\varphi}$ are trained, given as
\begin{subequations}
\begin{align}
  & {{f}_{\theta }}: {{\mathbf{r}}_{i}}\in {{\mathbb{R}}^{L\times 2}}\to {{\mathbf{y}}_{i}}\in {{\mathbb{R}}^{d}}, \\ 
 & {{g}_{\sigma }}: {{\mathbf{m}_{j}}}\in {{\mathbb{R}}^{a\times b}}\to {{\mathbf{n}_{j}}}\in {{\mathbb{R}}^{d}}, \\ 
 & {{h}_{\varphi }}: {{\mathbf{y}}_{i}}\to \widetilde{\mathbf{y}_{i}},
\end{align}
\end{subequations}
where ${\mathbf{m}}_{j}$ denotes the original $j$-th node feature of the KG, ${\mathbf{y}}_{i}$ and ${\mathbf{n}}_{j}$ denote the $i$-th signal features and the $j$-th semantic features, respectively, $\widetilde{\mathbf{y}_{i}}$ is the prediction result of ${\mathbf{r}}_{i}$, $\theta$, $\sigma$, $\varphi$ are the parameters of each module, $a$ and $b$ are the number of nodes in the KG and the dimension of initial node features. The total loss $l$ of this scheme can be calculated as
\begin{equation}
    l=\sum\limits_{i}{\sum\limits_{j}{\mathcal{L}(
    {{\mathbf{y}}_{i}},{{\mathbf{n}}_{j}},\widetilde{\mathbf{y}_{i}},p_{i,j}
    |\theta ,\sigma ,\varphi )}},\
\end{equation}
where $\mathcal{L}(\cdot)$ denotes the loss function of our proposed scheme which will be introduced, and $p_{i,j}$ denotes the probability that the $i$-th signal belongs to the $k$-th modulation. The loss function requires both signal and semantic features as inputs to update the networks using the gradient backpropagation. Thus, the knowledge encoded in the KG can impact the parameters of data-driven models.

\section{The Proposed KGAMC Scheme}
\begin{figure*}[!t]
\centerline{\includegraphics[width=\linewidth]{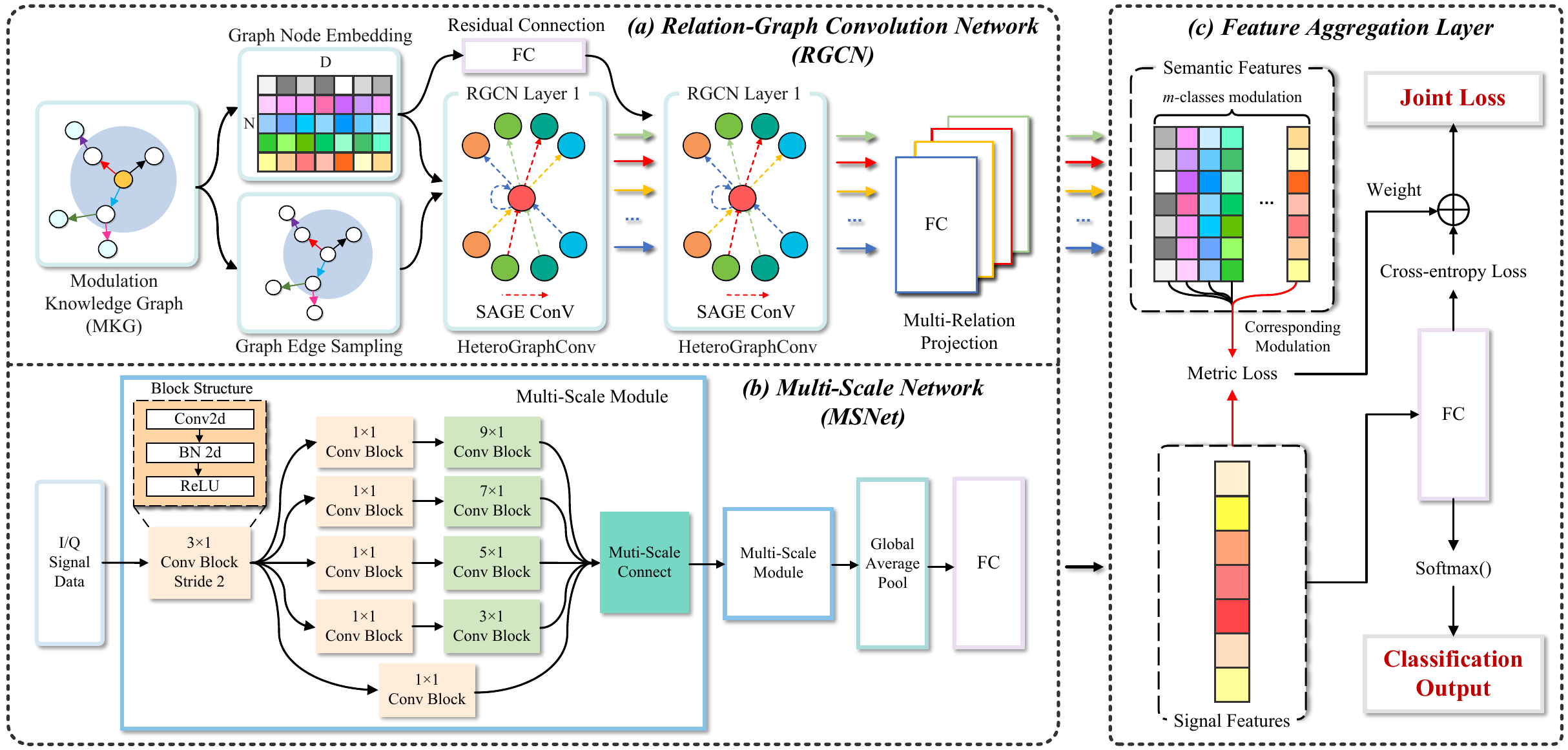}}
\caption{The scheme of KGAMC.}
\vspace{-0.5cm}
\label{fig1}
\end{figure*}
The proposed KGAMC scheme is illustrated in Fig.~\ref*{fig1}. The scheme has 3 parts with different functions. In Fig.~\ref*{fig1}(a), an improved relation-graph convolution network (RGCN) \cite{b12} is adopted to extract the semantic features of modulations by passing knowledge along the KG edges. In Fig.~\ref*{fig1}(b), an improved multi-scale network (MSNet) is utilized to extract the discriminative I/Q signal features on multiple time scales. In Fig.~\ref*{fig1}(c), a joint loss composed of the cross-entropy loss and the metric loss aims to exploit the semantic feature differences to aggregate the same signal features and separate the different signal features to achieve high modulation classification accuracy.
\subsection{MKG Construction and Embedding}
KG is a directed graph whose basic elements are the entity nodes and the relationship edges. The basic facts described in the KG are in the form of ordered triples, given as
\begin{equation}
\mathbf{T}=\left\{ (h,r,t)|h,t\in \mathbf{E},r\in \mathbf{R} \right\},\
\end{equation}
where $\mathbf{T}$ is the set of triples, $\mathbf{E}$ is the set of entities, $\mathbf{R}$ is the set of relations, $h$, $t$ and $r$ represent the head entity, tail entity, and their relations, respectively \cite{b13}.

In our proposed scheme, the MKG ought to expose the differences and similarities of information among various modulations in a comprehensive manner. Thus, the distinctive semantic features can guide the learning and updating process of the data-driven model, thereby enhancing the model interpretability. To obtain high accuracy in AMC, we construct a modulation-oriented radio KG, which combs the technical characteristics and application scenarios of modulations. Moreover, we note that the variability of knowledge structures has a vital impact on the embeddings of heterogeneous graphs. Thus, the MKG knowledge hierarchy is intentionally designed to ensure that the knowledge propagation over various edges is differentiated. To be specific, the relation types in the MKG are shown in Table~\ref*{tab1}, which present the ontology of the MKG. Subsequently, the relevant knowledge of modulations is collected, and the MKG is constructed by connecting the knowledge entities with relations to form triples based on the ontology.

\begin{table}[htbp]
\caption{Relation Types of MKG}
\begin{center}
\begin{tabular}{@{}>{\centering\arraybackslash}m{2.5cm}|>{\centering\arraybackslash}m{2.5cm}|>{\centering\arraybackslash}m{2.5cm}@{}}  
\toprule
\toprule
\textbf{Head Node Type} & \textbf{Relation} & \textbf{Tail Node Type} \\ \midrule
\toprule
modulationType     & possesses         & modualtionMethod   \\
base               & isBaseOf          & modualtionMethod   \\
bandwidthLevel     & hasBandwidthIn    & modualtionMethod   \\
situation          & adopts            & modualtionMethod   \\
modulationTheory   & includes          & modulationType     \\
carrierType        & isUsedIn          & modulationType     \\
dataType           & isModulatedBy     & modulationType     \\ \bottomrule
\end{tabular}
\vspace{-0.5cm}
\label{tab1}
\end{center}
\end{table}
After constructing the MKG, the initial feature embeddings of nodes that maintain the original semantic and structural information of the MKG are of vital importance due to the message-passing mechanism in the graph convolution process. Therefore, in our proposed KGAMC framework, the feature initialization of the nodes in the KG is mainly determined by the following attributes, including the number of first-order neighbors of the node, the number of the second-order neighbors, the out-degree, the in-degree, the node type, and the corresponding rows of the node in the adjacency matrix.
\vspace{-3pt}
\subsection{Models for Signal and Semantic Feature Extraction}
The core of achieving high accuracy in modulation classification lies in realizing effective feature extraction and differentiation on the blind signals. In the proposed framework, semantic features are used to aggregate the same class of signal features and separate the different classes of signal features. Consequently, the basis of the proposed KGAMC scheme is to extract stable and distinguishing semantic information on the MKG and map different signal features to the positions of corresponding modulations in the semantic space.

As shown in Fig.~\ref*{fig1}(a), an improved RGCN functioning as the semantic feature extractor for the MKG is designed. The improved RGCN consists of two HeteroGraphConv layers, a residual connection branch, and a multi-relation projection layer. Firstly, the whole-graph messages are gathered and extracted through two heterogeneous convolution layers, which contain graph convolution units for different relationships. And the GraphSAGE algorithm is utilized as the graph convolution units to learn a function that generates embeddings by sampling and aggregating features from a node local neighborhoods \cite{b14}, whose expression is
\begin{subequations}
\begin{align}
  & h_{N\left( i \right)}^{\left( l+1 \right)}=\operatorname{aggregate}\left( \left\{ \left. h_{j}^{l},\forall j\in N\left( i \right) \right\} \right. \right), \\ 
 & h_{i}^{\left( l+1 \right)}=\operatorname{norm}\left( \sigma \left( \mathbf{W}\cdot\operatorname{concat}\left( h_{j}^{l},h_{N\left( i \right)}^{\left( l+1 \right)} \right) \right) \right),
\end{align}
\end{subequations}
where $h_{j}^{l}$ is the feature of node $j$ in the $l$th iteration, $N(i)$ is the neighbor node set of node $i$ and $\mathbf{W}$ is the parameter matrix. Secondly, a FC layer works as a residual connection branch to avoid the oblivion of original information and to accelerate the RGCN convergence. Finally, after accomplishing the semantic features spreading by 2 RCGN layers, a multi-relation projection layer consisting of several FC layers is used to project the embeddings into semantic space that enhances the quality of embeddings for the introduction of learnable nonlinear transformation. Additionally, the activation function used in the RGCN between the layers is the leaky rectified linear function (Leaky-ReLU) to introduce non-linearity into the RGCN and capture more information over the ReLU.

Simultaneously, an improved MSNet is employed as the signal features extractor shown in Fig.~\ref*{fig1}(b), which is composed of two multi-scale blocks, a global average pooling (GAP) and a FC layer. To obtain various time-scale signal features, two multi-scale blocks are adopted. The multi-scale blocks are formed by a 3×1 convolution block with a stride of 2 to reduce feature dimension and five convolution branches to extract different level information. The convolution branch consists of a 1×1 convolution block ahead and a convolution block with customized kernel size in series. The kernel size is 1×1, 3×1, 5×1, 7×1 and 9×1, respectively. Through these branches, the MSNet is capable to learn the similarities of the same modulation class signals in different sequential lengths. Afterwards, a global average pooling operation is used to merge multi-scale features. GAP is beneficial for maintaining the original spatial features and reducing the over-fitting problem since there is no parameter to be optimized. Lastly, the fused features are transferred into the semantic space by a FC layer as well, enabling the MSNet to learn the projection relation between the signal data space and semantic space.

Different from other data-and-knowledge dual-driven schemes, the distinctive semantic differences in the MKG exist between different modulations that create a larger decision interval, since the knowledge output by the RGCN illumes the learning direction of the data-oriented MSNet to map signal features into semantic space. This implies that the RGCN and the MSNet are trained in parallel only in the model optimization phase. However, during the model reasoning phase, only the well-trained MSNet is exploited to analyze and classify the signals. This approach makes the KGAMC scheme easier to deploy on miniaturized distributed devices. 

\subsection{Feature Aggregation Loss Function and Training Strategy}
Unlike general classification tasks, where the models are guided to output the correct classification probabilities only by the cross-entropy loss function, in order to fully exploit the knowledge, the KGAMC requires  additional loss items for the KG to guide the training process of the networks. Inspired by the N-pair loss in the contrastive learning \cite{b15}, a metric-based loss function is proposed to exploit the semantic features of knowledge to aggregate data features and reduce the inter-class distance.

Fig.~\ref*{fig1}(c) describes the feature aggregation process. Specifically, after the RGCN and the MSNet independently extract and distinguish the features of the corresponding domains, the cosine similarity  ${y}_{cik}$ between each signal feature $\mathbf{x}_{i}\in{\mathbb{R}^{d}}$ and each semantic feature $\mathbf{x}_{sk}\in{\mathbb{R}^{d}}$ are calculated. Then, ${y}_{cik}$ activated by the $softmax$ activation function is considered as the predicted probability. Lastly, the cross-entropy loss is deployed to measure the N-pair loss $\mathcal{L}_{\mathrm{npair}}$ between the predicted probability and the true probability. The N-pair loss $\mathcal{L}_{\mathrm{npair}}$ is given as
\begin{subequations}
\begin{align}
  {{y}_{\operatorname{c},\ i,k}}&=\frac{{{\mathbf{x}}_{i}} {{\mathbf{x}}_{sk}}}{\left| {{\mathbf{x}}_{i}} \right| \left| {{\mathbf{x}}_{sk}} \right|}, \\ 
 \mathcal{L}_{\mathrm{npair}}&=-\frac{1}{N}\sum\limits_{i=1}^{N}{\sum\limits_{k=1}^{M}{p({{y}_{\operatorname{c},\ i,k}})}\log (\operatorname{softmax}({{y}_{\operatorname{c},\ i,k}}))},
\end{align}
\end{subequations}
where $M$ is the number of the modulations needed to be classified, $N$ is the batch size, and $p({y}_{cik})$ represents the true probability belonging to the $i$-th signal feature of the $k$-th modulation. The advantage of the cross-entropy form is to gently propagate the gradient, and the cosine similarity reduces the impact of amplitude differences in the same class.

Moreover, to enable the MSNet to learn the discriminative signal features, the angles between different semantic features should be larger than 90°. For instance, the traditional classification space can be considered as a $M$-dimensional space, and the one-label classification task is making each modulations’s true label on different orthogonal axis, which is the one-hot encoding. However, in the $\mathbb{R}^{d}$ semantic space of dimension $d$ ($d>M$), it is capable to create larger decision space in the hypersphere for $M$ categories. Hence, the penalty term $\mathcal{L}_{\mathrm{p}}$ is added to the metric loss function to enlarge the angles, and it can be expressed as
\begin{subequations}
\begin{align}
  & {{y}_{p}}=\frac{2}{M(M-1)}\sum\limits_{k=1}^{M}{\sum\limits_{l=1}^{M}{\frac{{{\mathbf{x}}_{sl}} {{\mathbf{x}}_{sk}}}{\left| {{\mathbf{x}}_{sl}} \right| \left| {{\mathbf{x}}_{sk}} \right|}}},l\ne k, \\ 
 & \mathcal{L}_{\mathrm{p}}=\max (0,{{y}_{{p}}}),  
\end{align}
\end{subequations}
where ${y}_{{p}}$ is the average cosine similarity between each semantic features of the modulations. 
\begin{figure}[htbp]
\centerline{\includegraphics[width = \linewidth]{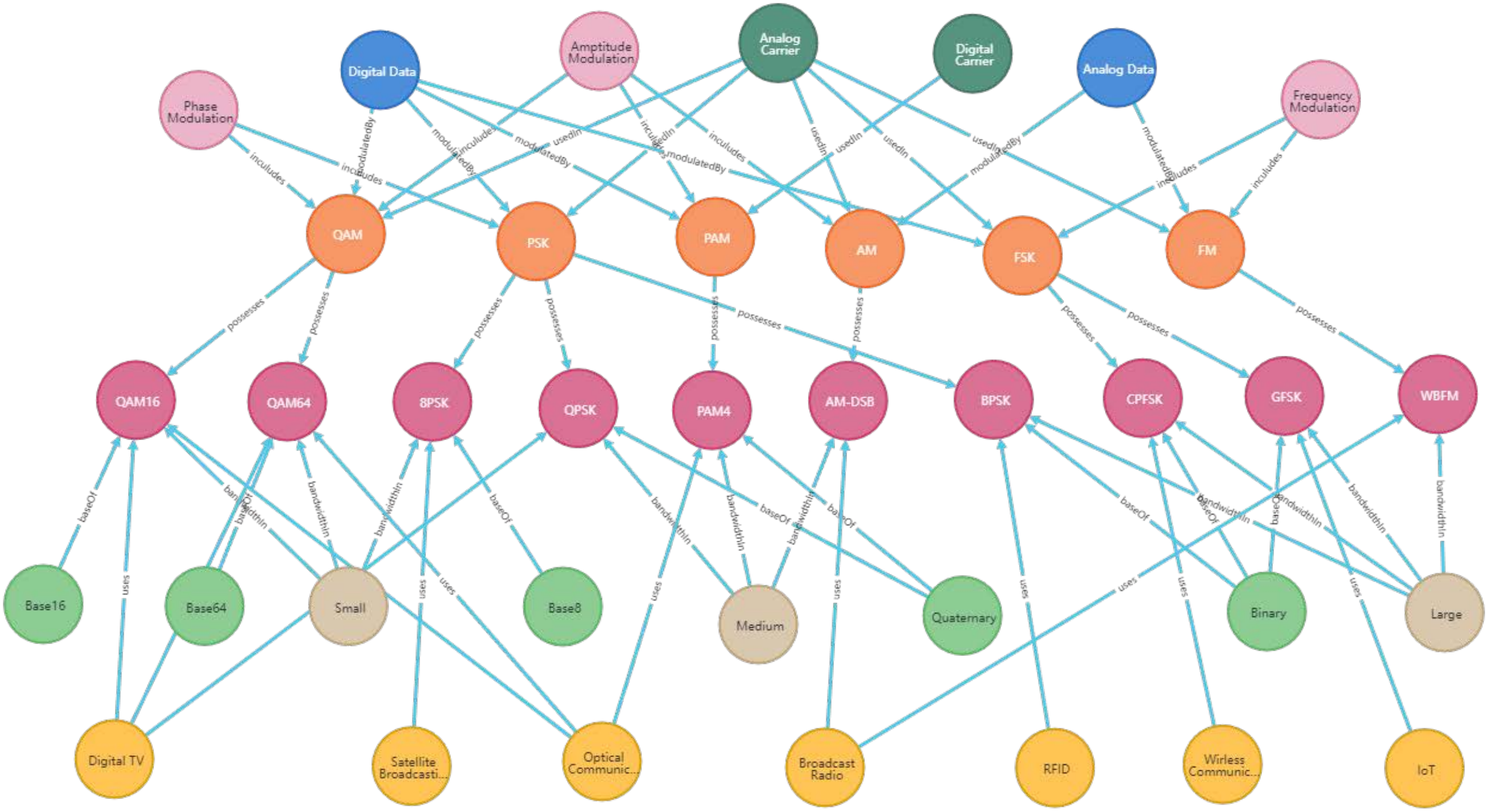}}
\caption{The MKG constructed for RadioML 2016.10b.}
\vspace{-0.5cm}
\label{fig2}
\end{figure}
Besides, the signal features are also input into a FC classifier and a cross-entropy loss $\mathcal{L}_{\mathrm{ce}}$ is calculated. The joint loss of KGAMC $\mathcal{L}$ is formulated as
\begin{subequations}
\begin{align}
 \mathcal{L}_{\mathrm{ce}} & =-\frac{1}{N}\sum\limits_{i=1}^{N}{\sum\limits_{k=1}^{M}{p({{y}_{ik}})}\log (\operatorname{softmax}({{y}_{ik}}))}, \\ 
 \mathcal{L} &=\mathcal{L}_{\mathrm{ce}}+\lambda (\mathcal{L}_{\mathrm{npair}}+\mathcal{L}_{\mathrm{p}}), 
\end{align}
\end{subequations}
where ${y}_{ik}$ is the probability that the $i$-th signal belongs to the $k$-th modulation, and the weight $\lambda$ of the metric loss is used to balance the learning rate of two different loss functions.

Through the metric loss, the models learn how to form a semantic space of the MKG at once and how to derive the discriminating signal features by projecting the signal features to the semantic space. Notably, in order to adapt the learning progress of the model and ensure that the semantic features in the semantic space are as stable as anchors during the training process, the learning rate of the parameters of the RGCN should be several orders of magnitude lower relative to that of MSNet. The stable semantic features of modulations accelerate the aggregation of the same class signal features, making the joint loss more efficient for KGAMC.

\footnotetext[1]{https://www.deepsig.ai/datasets/}
\section{Simulation Results}
In this section, the effectiveness of our proposed KGAMC is verified by presenting simulation results obtained from RadioML 2016.10b dataset \cite{b4}, which is presented on the website\textsuperscript{1}. The dataset  encompasses 8 digital and 2 analog modulations spanning a SNR range from -20dB to 18dB with a step of 2dB. Each type of modulations contains 6000 samples per SNR for a total of 1,200,000 signal samples. The dimension of the sampled I/Q signals is 2×128. The dataset is divided into the training set and testing set at a ratio of 8:2. Based on the discussion in section III-A, the MKG should be constructed with the general knowledge of modulations in the dataset under the restriction of the ontology, and the overview of the MKG constructed for this dataset is shown in Fig.~\ref*{fig2}.
\begin{figure}[htbp]
\centerline{\includegraphics[width = \linewidth]{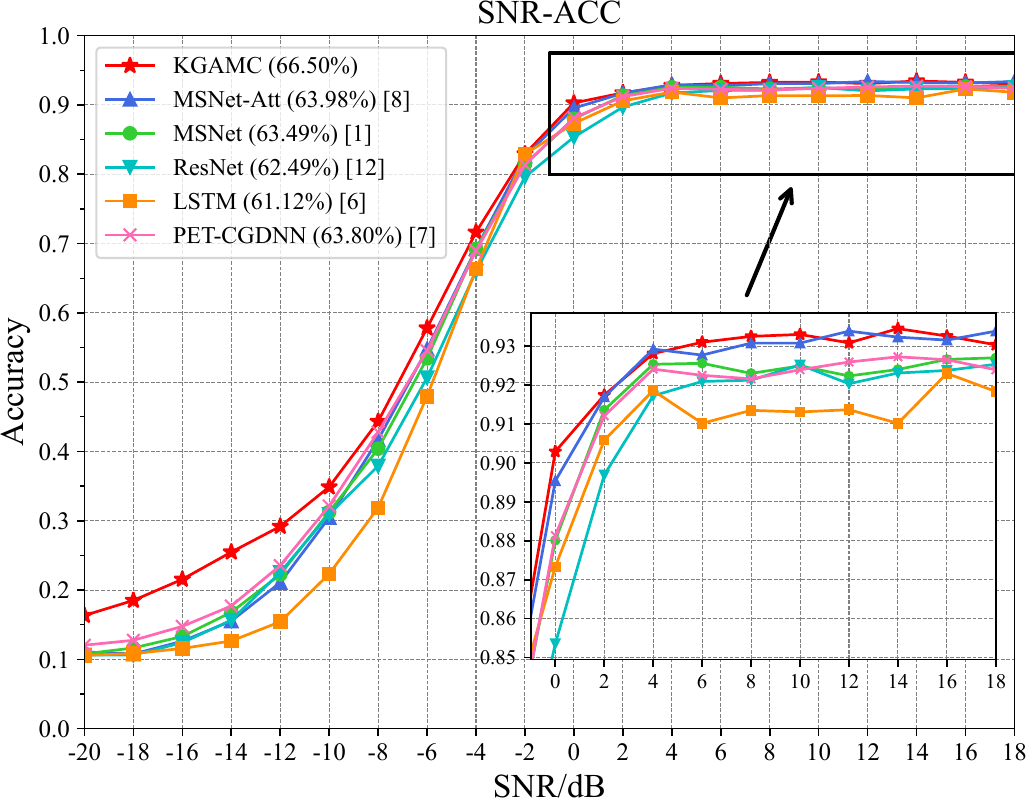}}
\setlength{\abovecaptionskip}{0.cm}
\setlength{\belowcaptionskip}{-0.cm}
\caption{Comparison of the classification performance among MSNet-Att\cite{b7}, MSNet\cite{b16}, ResNet\cite{b11}, LSTM\cite{b5}, PET-CGDNN\cite{b6} and our proposed KGAMC.}
\vspace{-0.5cm}
\label{fig3}
\end{figure}

The networks of KGAMC are built on a Linux server with the PyTorch platform and Deep Graph Library (DGL) \cite{b17}. Since the HeteroGraphConv layers are comprised by the DGL neural network modules, the MKG should be transformed into a computable DGL heterography object first. The Adam optimizer is adopted to update the models, and the weight decay is 0.0005. The initial learning rate is set at 0.001 for the MSNet and 0.000001 for the RGCN, and a StepLR scheduler is utilized to reduce 20\% learning rate every 5 epochs. The number of training epochs is set as 80 with a batch size $N=1024$. For the hyperparameters involved in the experiment, we choose the feature dimension $d=128$ and the weight $\lambda=0.2$ to obtain the best performance.

Sseveral crucial AMC models are implemented to provide a benchmark comparison with our proposed KGAMC, including MSNet-Att \cite{b7}, MSNet \cite{b16}, ResNet \cite{b11}, LSTM \cite{b5}, PET-CGDNN \cite{b6}. 
Fig.~\ref*{fig3} shows the classification accuracy of each modulation in the testing set at varying SNRs. Additionally, the legends in Fig.~\ref*{fig3} highlight the overall average  accuracy of all modulations at all SNRs. It is evident that our proposed KGAMC scheme is superior to other baseline models. The KGAMC achieves the highest overall accuracy of 66.50\% compared to others, exceeding the MSNet-Att with second highest accuracy by 2.02\%, and it also outperforms other compared methods in the SNR range of -20dB to 0dB. The accuracy of an AMC model at -20dB for a classification task with 10 modulations is typically around 10\%, while KGAMC can reach an accuracy rate of 16.33\%. Simultaneously, the KGAMC continues to demonstrate top-tier classification performance for modulations with SNR exceeding 0dB that demonstrates the effectiveness of our scheme. In this sense, it confirms that it is beneficial to project the signal features into the semantic space and the MKG provides MSNet with a clearer insight into the characteristics of different modulations. The considerable boost in classification precision at low SNR holds a vital importance for AMC schemes to adapt to the complex wireless communication environment. 
\begin{figure}[htbp]
	\centerline{  
	\subfigbottomskip=0pt 
	\subfigcapskip=-5pt 
	\subfigure[]{
		\includegraphics[width=0.5\linewidth]{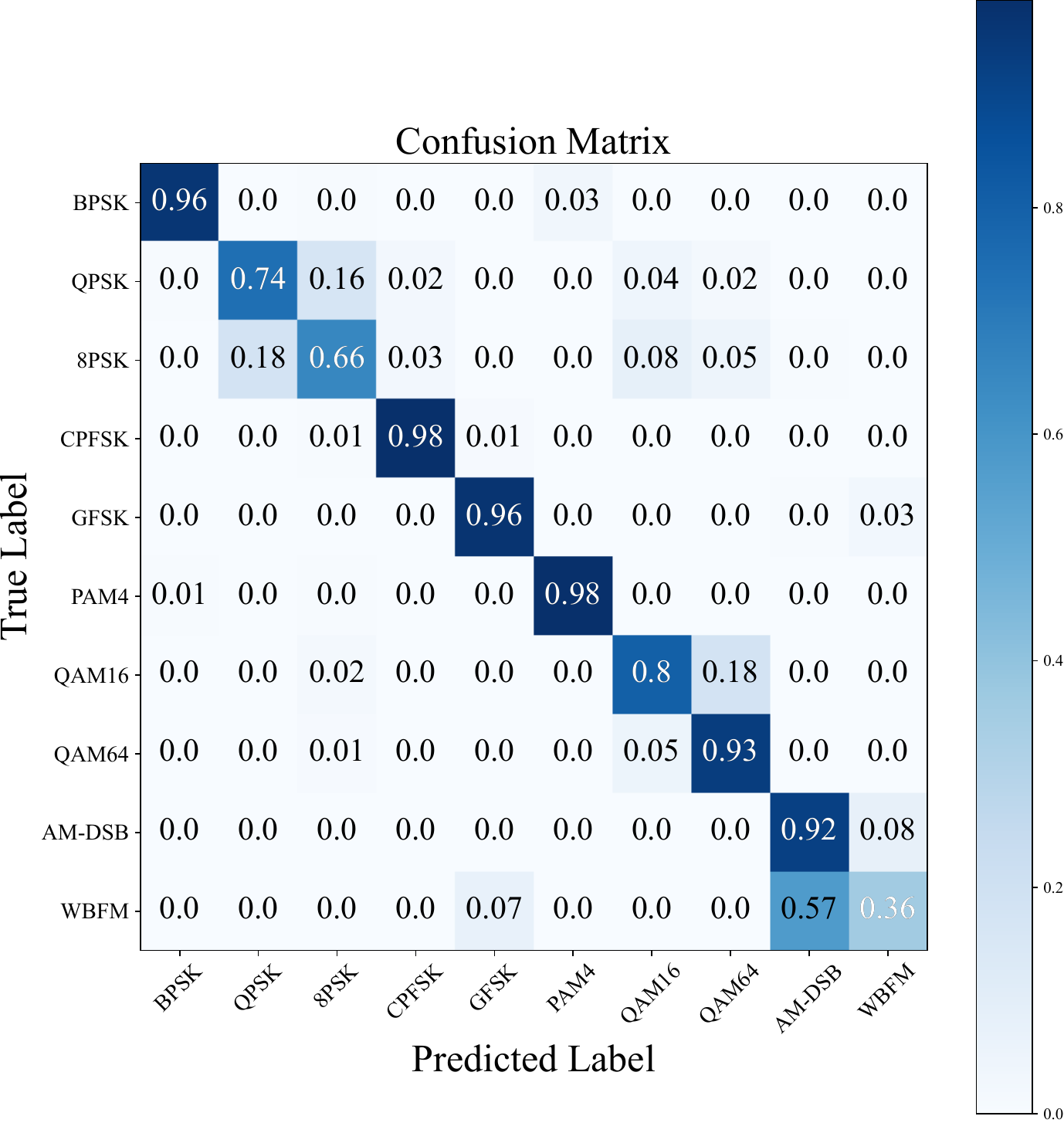}}
	\subfigure[]{
		\includegraphics[width=0.5\linewidth]{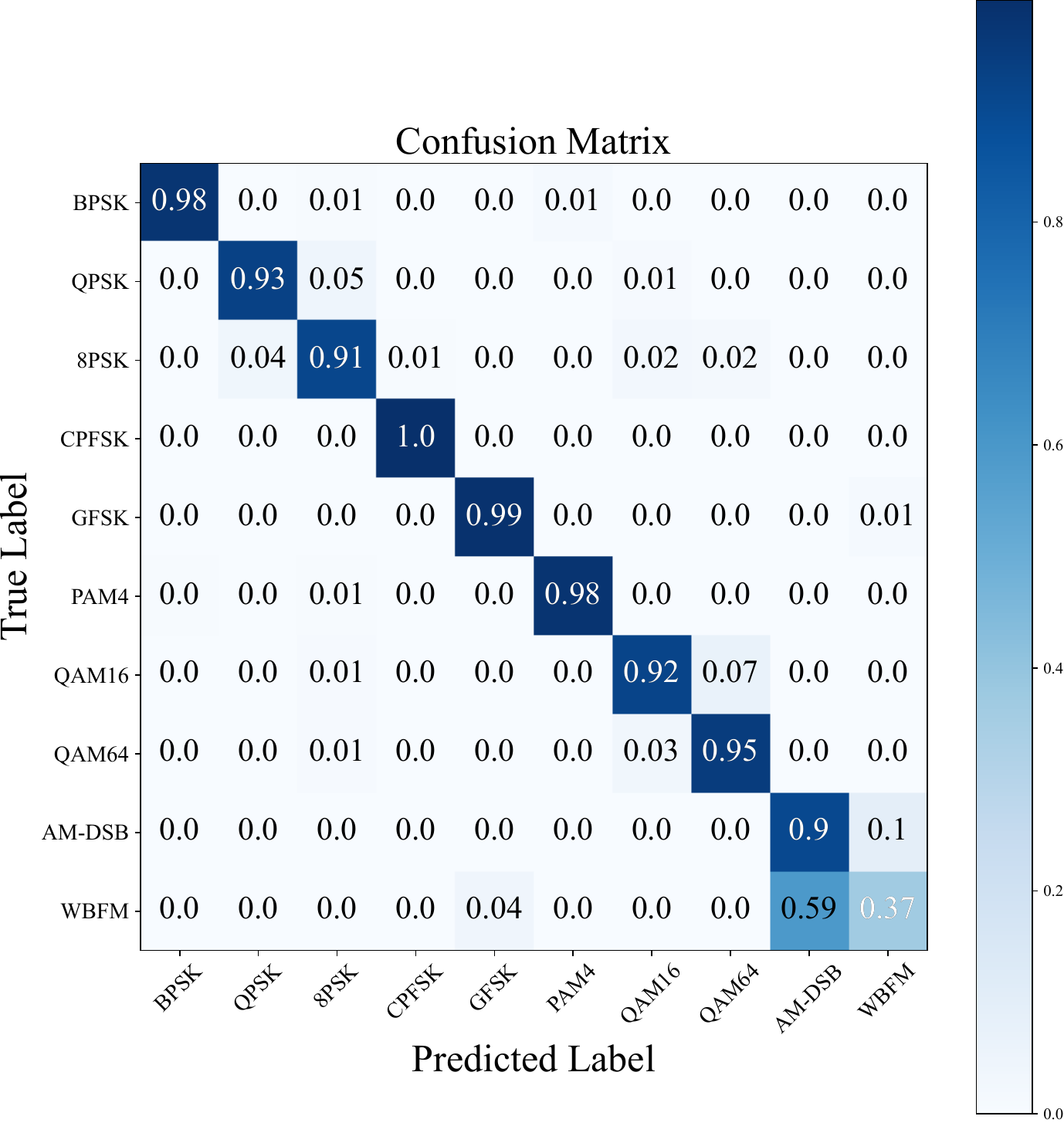}}}
	\caption{Comparison of confusion matrix at 0dB, (a) MSNet \cite{b16}, and (b) Our proposed KGAMC.}
    \label{fig4}
\end{figure}
\begin{figure}[htbp]
	\centerline{  
	\subfigbottomskip=-5pt 
	\subfigcapskip=-5pt 
	\subfigure[]{
		\includegraphics[width=0.48\linewidth]{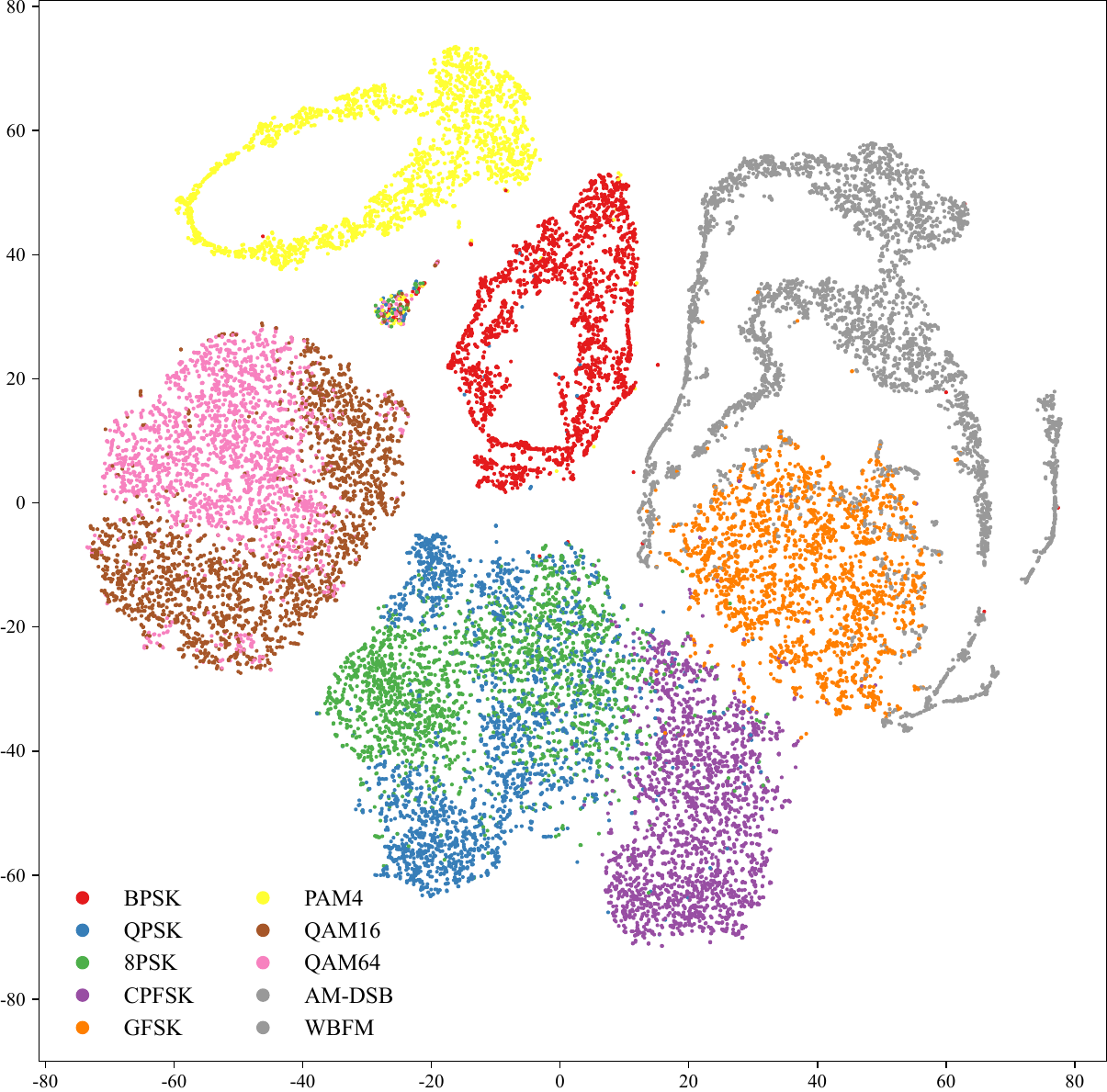}}
	\subfigure[]{
		\includegraphics[width=0.48\linewidth]{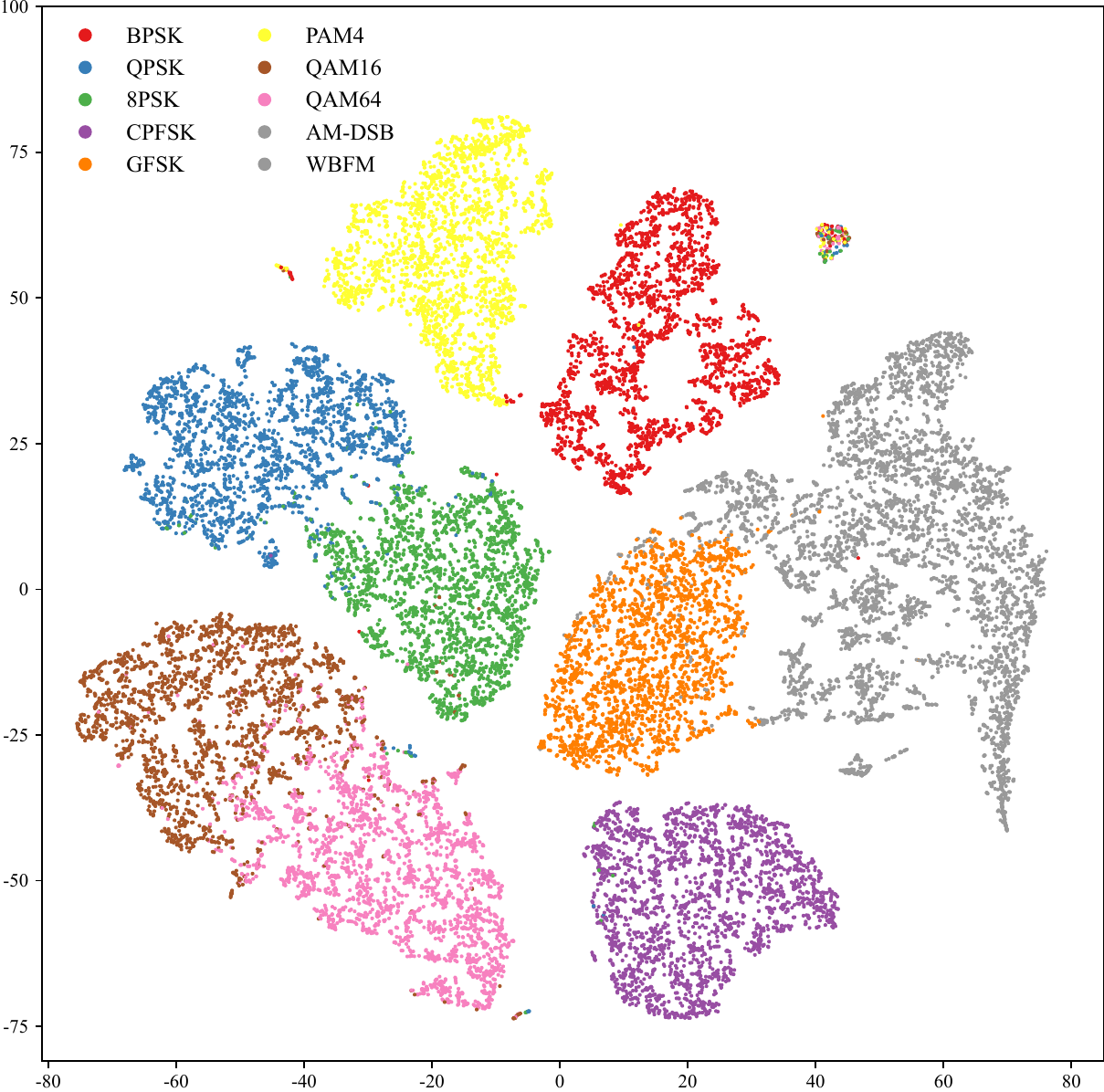}}}
	\caption{The t-SNE \cite{b18} visualization of signal features extracted by (a) MSNet \cite{b16}, and (b) Our proposed KGAMC.}
    \label{fig5}
\vspace{-0.2cm}
\end{figure}

In Fig.~\ref*{fig4}, a comparison of the confusion matrices at 0dB of MSNet and the KGAMC is displayed. It is clear that the misclassification between QPSK and 8PSK, QAM16 and QAM64 is alleviated with the help of the MKG. Only the AM-DSB and WBFM achieve the worst performance in both schemes. As stated in \cite{b3}, this challenge arises from the relatively small observation window employed during the dataset's analog signal generation process, indicating that the models have difficulty in using identical network parameters to classify the digital and analog signals.

To further evaluate the effectiveness of the metric loss and the MKG in feature aggregation, we visualize the extracted signal features at 0dB and 2dB by converting them into 2-dimensional scatter map with t-SNE \cite{b18} operation in Fig.~\ref*{fig5}. The scatter map generated by t-SNE operation illustrates relative distance between the features. The signal features of our proposed KGAMC are considerably more aggregated within the inter-class and more separated between different classes compared to the MSNet alone. The signal features of high-order modulations including QPSK, 8PSK, QAM16 and QAM64 are more discriminative than those extracted by MSNet. Hence, we can draw the conclusion that our proposed KGAMC offers an attractive solution to mitigate the challenge of high-order modulations in AMC.

\section{Conclusion}
A novel KG driven AMC scheme, named KGAMC, was proposed by aggregating the signal features with semantic features of the MKG. The MKG is constructed with the fundamental knowledge of modulations. The differences between the modulations in knowledge domain were extracted by the improved RGCN with a residual branch, and enhanced the performance of MSNet in aggregating and separating the signal features by adopting a designed metric loss, which reduced the distance between signal features and semantic features of the same modulation. Simulation results demonstrated that our proposed KGAMC scheme was superior to other benchmark schemes in terms of classification accuracy and feature attribution, especially in the low SNR range. The KG driven method reveals encouraging prospects in improving the  performance and interpretability of data-driven models, and is worth an in-depth investigation on addressing other AMC challenges.

\bibliographystyle{IEEEtran}
\bibliography{IEEEabrv,reference}

\end{document}